\documentclass[preprint2,11pt]{aastex}
\usepackage{graphicx}

\usepackage[dvips]{geometry}
\newcommand{\microns}{\micron}

\newcommand{\eg}{{e.g.}}
\newcommand{\etal}{{et~al.}}
\newcommand{\decon}{W4$_{\mathrm{decon}}$}

\slugcomment{}
\shorttitle{The Infrared Rings of NGC~1514}
\shortauthors{Ressler \etal}

\begin{document}

\title{The Discovery of Infrared Rings in the Planetary Nebula NGC~1514 During the WISE All-Sky Survey}

\author{Michael E. Ressler}
\affil{Jet Propulsion Laboratory, California Institute of Technology,\\
4800 Oak Grove Drive, Pasadena, CA 91109}
\email{Michael.E.Ressler@jpl.nasa.gov}

\author{Martin Cohen}
\affil{Monterey Institute for Research in Astronomy\\
200 8th Street, Marina, CA 93933}

\author{Stefanie Wachter, D. W. Hoard}
\affil{Spitzer Science Center, California Institute of Technology,\\
1200 E California Blvd, Pasadena, CA 91125}

\author{Amy K. Mainzer}
\affil{Jet Propulsion Laboratory, California Institute of Technology,\\
4800 Oak Grove Drive, Pasadena, CA 91109}

\and{}

\author{Edward L. Wright}
\affil{UCLA Astronomy\\
PO Box 951547, Los Angeles, CA 90095}

\begin{abstract}
We report the discovery of a pair of infrared, axisymmetric rings in
the planetary nebula NGC~1514 during the course of the WISE all-sky
mid-infrared survey. Similar structures are seen at visible
wavelengths in objects such as the ``Engraved Hourglass Nebula''
(MyCn~18) and the ``Southern Crab Nebula'' (Hen~2-104). However, in
NGC~1514 we see only a single pair of rings and they are easily
observed only in the mid-infrared. These rings are roughly 0.2 pc in
diameter, are separated by 0.05 pc, and are dominated by dust emission with a
characteristic temperature of 160 K. We compare the morphology and
color of the rings to the other nebular structures seen at visible,
far-infrared, and radio wavelengths, and close with a discussion of a
physical model and formation scenario for NGC~1514.
\end{abstract}


\section{Introduction}

In 1790, William Herschel found ``a most singular phaenomenon! A star
of about the 8th magnitude, with a faint luminous atmosphere, of a
circular form, and of about 3\arcmin\ in diameter.'' This object, now
known to us as NGC~1514, with its central star inseparable from a
``shining fluid'', convinced him that not all the nebulae he studied
could be resolved into clusters of stars as he had thought
\citep{herschel1791}. Thus began over 200 years of study of NGC~1514,
though it is still classified as the round or elliptical planetary
nebula that Herschel saw.

In modern terms, NGC~1514 is a moderately high excitation planetary
nebula (PN) in Taurus ($\alpha$ = 04$^{\mathrm h}$ 09$^{\mathrm
  m}$ 16\fs990, $\delta$ = +30\degr{} 46\arcmin{} 33\farcs44, J2000). The
optically visible central source (CSPN) of NGC~1514 is unusual in that
its spectral type is too late (A0) to produce the observed nebula.
\citet{kohoutek1967} and \citet{kohoutek1967b} developed the notion of a binary
central source, and concluded that an $M_V = -1.4$ A0III giant and a
sub-luminous, dwarf O star would reproduce the observed UV spectrum.
By contrast, \citet{greenstein1972} concluded through the use of
higher spectral resolution spectrophotometry that the visible star is
a horizontal branch A star. This star is slightly cooler and has a much
lower absolute magnitude than an A0III giant. He finds that the CSPN
is better represented by an $M_V = +0.8$ A star of $\lesssim 10000$~K,
and an sdO star with a temperature of 100,000~K and $M_V = +2.8$.
Later, \citet{feibelman1997} built on earlier IUE observations by
\citet{seaton1980} to show that the CSPN could be represented by an
early A star of $\sim 9000$~K along with a $\gtrsim$ 60,000~K
sub\-dwarf; he also found that the CSPN's ultraviolet flux varied by more than a
factor of two on timescales of a year to a decade.

\begin{figure}[!ht]
\begin{center}
\includegraphics[width=1\columnwidth]{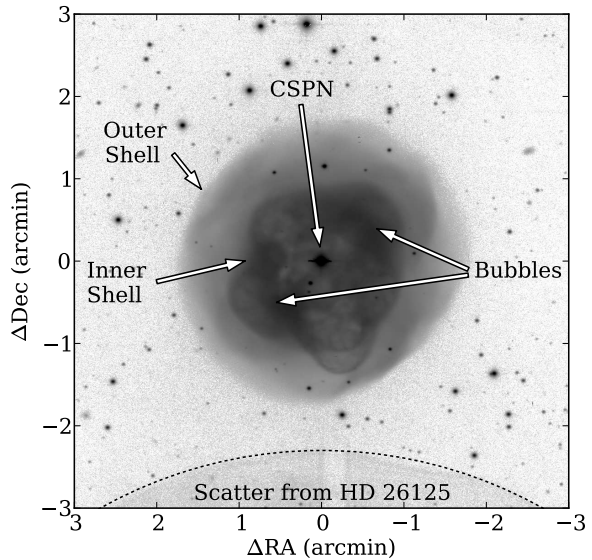}
\end{center}
\caption{Labeled, logarithmically-scaled, H$\alpha$ image of NGC~1514.
  (FITS data kindly supplied by R. Corradi.)\label{fig:guide}}
\end{figure}

NGC~1514 itself is typically catalogued as a round or slightly elliptical
nebula, with an amorphous appearance \citep[\eg][]{balick1987}.
\citet{kohoutek1968} describes the nebula as having an inner shell (or
main body) with a diameter of $\sim 136$\arcsec, and an outer,
homogeneous, spherical layer of $\sim 206$\arcsec. He interprets
condensations within the inner part of the nebula as a toroid with the
axis of symmetry at a position angle of 35\arcdeg. \citet{chu1987}
classify it as a double shell PN, 132\arcsec{} and 183\arcsec\ in
diameter. Figure~\ref{fig:guide} illustrates the optical appearance
and uses their nomenclature (which we adopt) to highlight the various features.

\citet{hajian1997} describe NGC~1514 as a ``lumpy nebula composed of
numerous small bubbles'' based on their deep [\ion{O}{3}] image. They
interpret the bubbles at the edge of the inner shell to be sweeping up
outer shell (``halo'') material which is providing pressure
confinement. They confirm a somewhat filamentary structure in the
outer shell that had been reported earlier by \citet{chu1987}.
Unfortunately, NGC~1514 does not have an observable halo of the type
needed for the timescale correlation method for determining the distance
that is the focus of their paper.

Perhaps the spatiokinematic study by \citet{muthu2003} is the most
in-depth look at the formation and structure of NGC~1514. They used an
imaging Fabry-P\'erot spectrometer to produce velocity-resolved maps
of the 5007 \AA\ [\ion{O}{3}] line. From modeling the double- and
triple-component line profiles seen in these maps, they deduced that
NGC~1514 is a generally ellipsoidal shell nebula, but that it also has
two ``blobs'' or bubbles of material emanating from the center. The southeast
bubble is slightly blue-shifted, and the northwest blob is red-shifted,
thus defining a ``polar'' axis. The bubbles themselves are 
uncollimated in that there are large velocity dispersions across each
of the bubbles; this is in contrast to other highly collimated, bipolar
flows seen in many other PNe. Finally, they model the overall velocity
structure of the nebula and derive peak expansion velocity of
23~km~s$^{-1}$ near the center, falling off with radial distance as
expected from projection effects along the line of sight.

\section{Observations}

The Wide-field Infrared Survey Explorer (WISE, \citealt{wright}) is a
NASA Mid-class Explorer mission that is surveying the entire sky at
wavelengths of 3.4, 4.6, 12, and 22 \microns{} (W1 through W4,
respectively). The point source sensitivity (defined as a 5$\sigma$
detection with eight repeat exposures per source) is currently
estimated to be 0.08, 0.11, 1, and 6~mJy in these passbands,
respectively. WISE surveys the sky in 47\arcmin{} wide strips with a
significant overlap between each strip that yields the eight or more
coverages (depending on ecliptic latitude). NGC~1514 was thus observed
twelve independent times on 2010 Feb 18-19 UT for a total of 106
seconds of integration time in each of the four bands; NGC~1514 was
easily detected in all four bands.

The data presented here were processed with the ``first-pass'' version
of the WISE pipeline that utilized preliminary reduction algorithms
and calibrations derived from early survey data. The twelve individual
fields have been corrected for detector artifacts, optical
distortions, and latent images. They were then combined into ``ops
coadds'' with pixel outlier rejection to suppress cosmic rays and
transient noise events, and have been resampled from the native plate
scales of 2.75 (W1, W2, and W3) and 5.5 (W4) arcsec pixel$^{-1}$ to
the WISE image atlas scale of 1.375 arcsec pixel$^{-1}$. The
photometry has been calibrated using WISE measurements of blue
(stellar) calibrators; calibration using red (ULIRG) calibrators
currently differs by several percent, so that the absolute fluxes are
currently uncertain by $\pm$10\%. (See \citealt{wright} for further
discussion.) No color corrections have been applied to the photometry
reported here. For our purposes, the correction terms for stellar
spectra and blackbody continua and for all sources in the W4 band are
essentially zero. In W3, the corrections do not exceed 15\% until
objects are cooler than $\sim$ 160 K. For W2, the equivalent
temperature is $\sim$ 250 K, and for W1, $\sim$ 400 K.

The W4 data were also reprocessed using the deconvolution capabilities
of AWAIC, a coaddition and deconvolution package developed for WISE,
but which can be used with generic astronomical data sets
\citep{masci2009}. The deconvolution algorithm ``HiRes'' is based on
the Maximum Correlation Method, which was developed to enhance the
scientific return from the IRAS survey. W4 data covering a 24~arc\-minute
field was passed through just three iterations of HiRes, enough so that the
model image has an apparent spatial resolution equivalent to the
native W3 resolution. The resulting image, denoted as \decon, is used
only in morphological discussions; all flux measurements are based on
the W4 ops coadd image.

\section{Results}

\begin{figure*}
\begin{center}
\begin{tabular}{@{} c c @{}}
\includegraphics[width=0.48\textwidth]{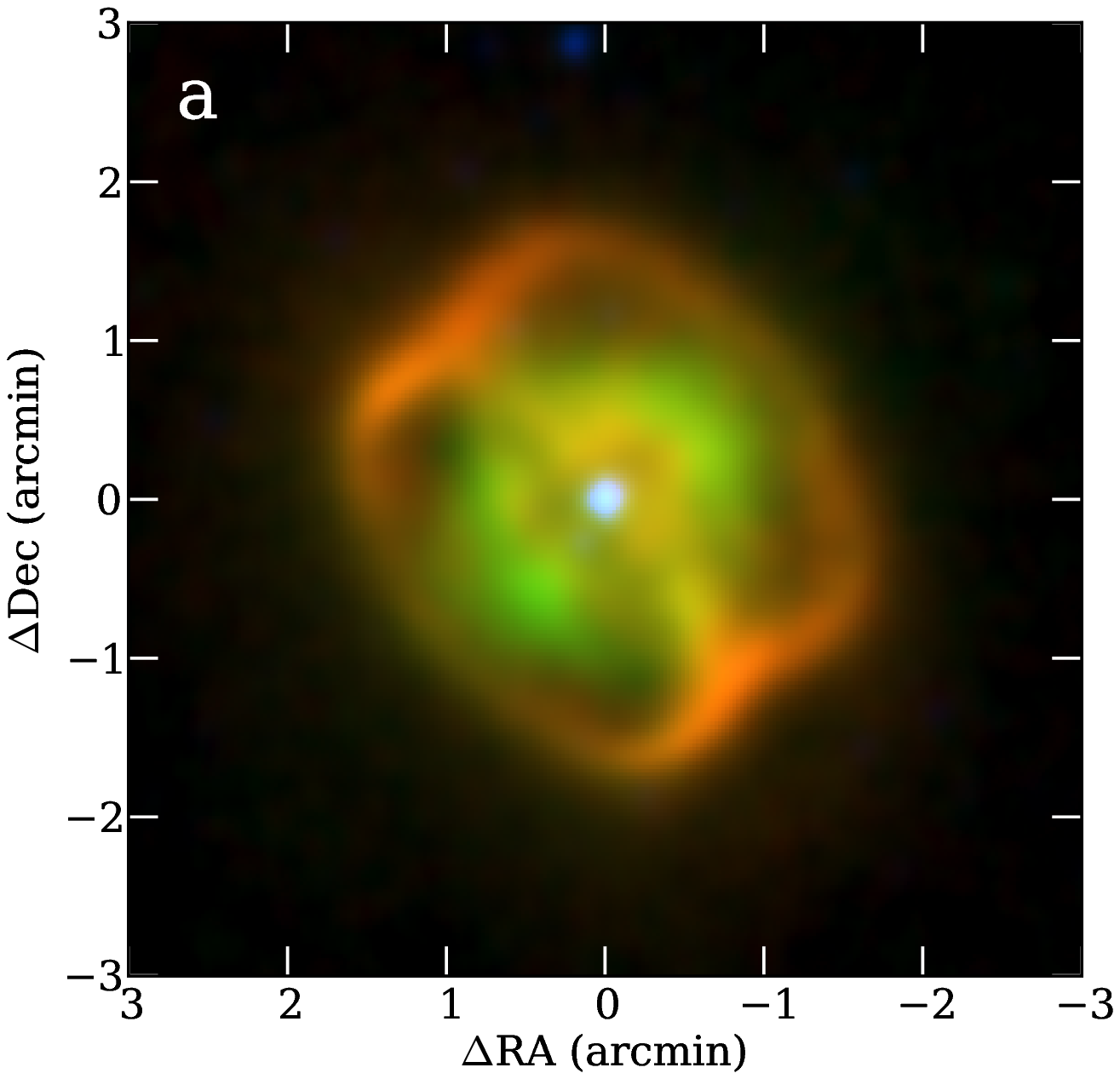} &
\includegraphics[width=0.48\textwidth]{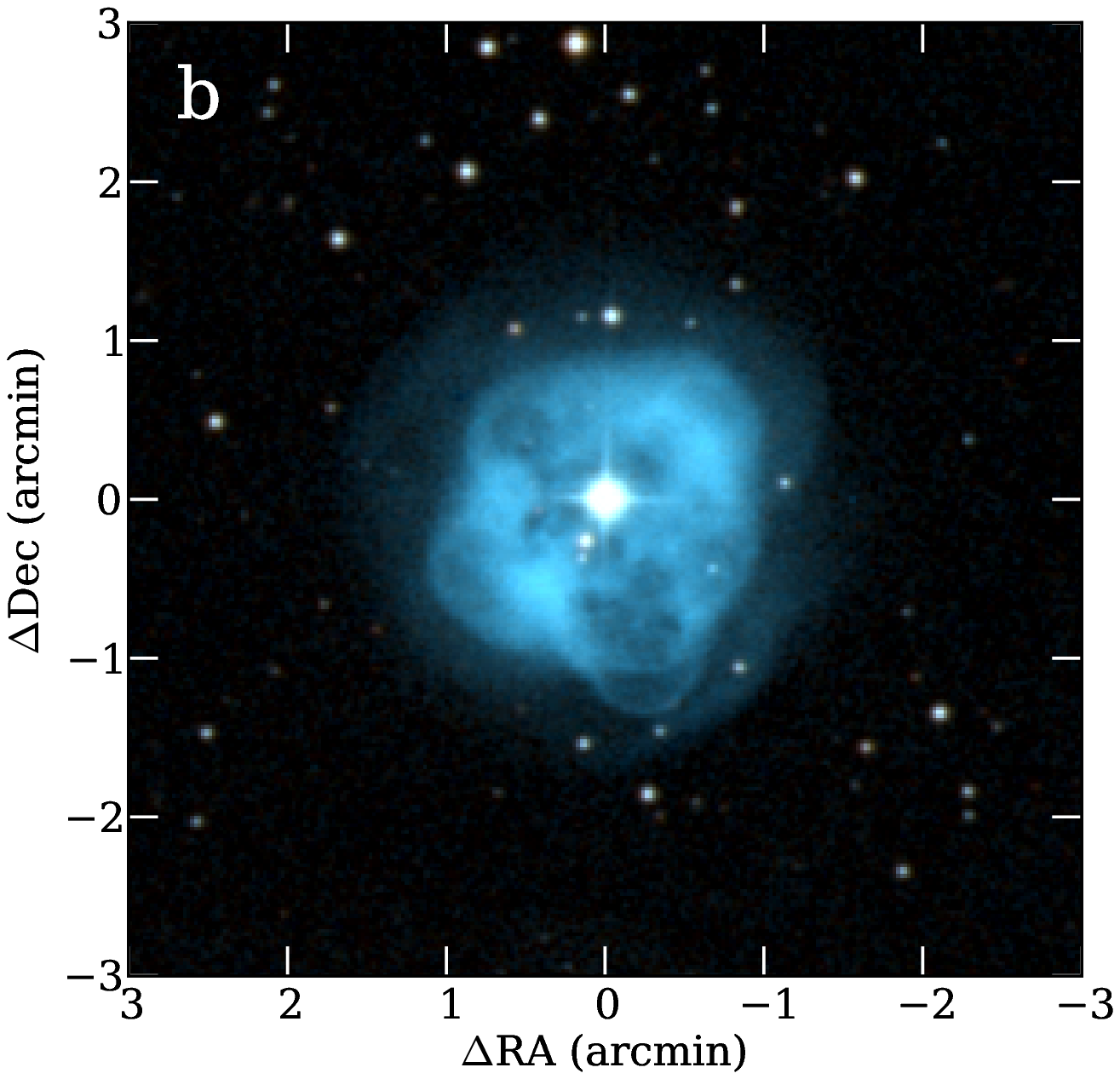} \\
\includegraphics[width=0.48\textwidth]{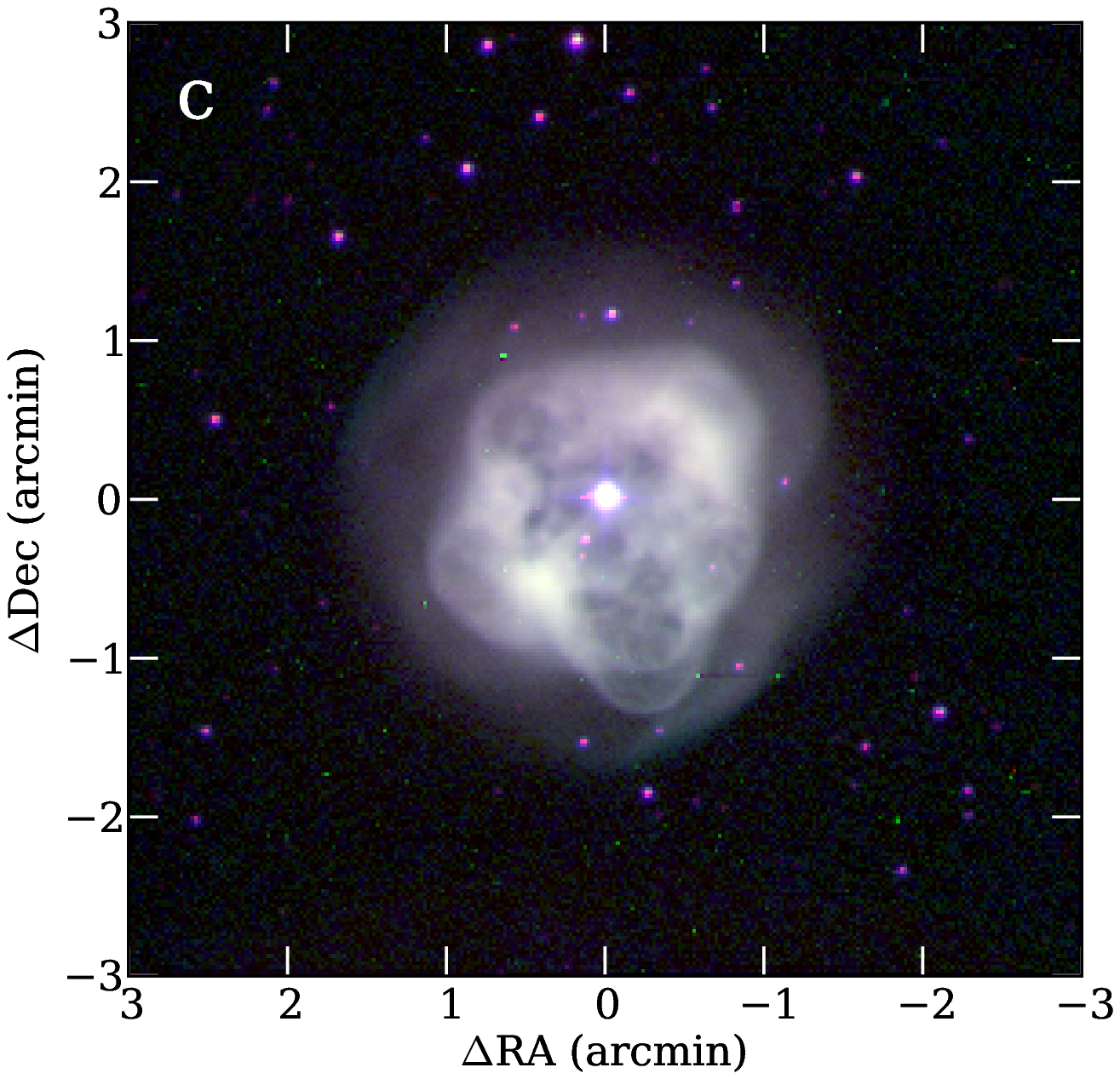} &
\includegraphics[width=0.48\textwidth]{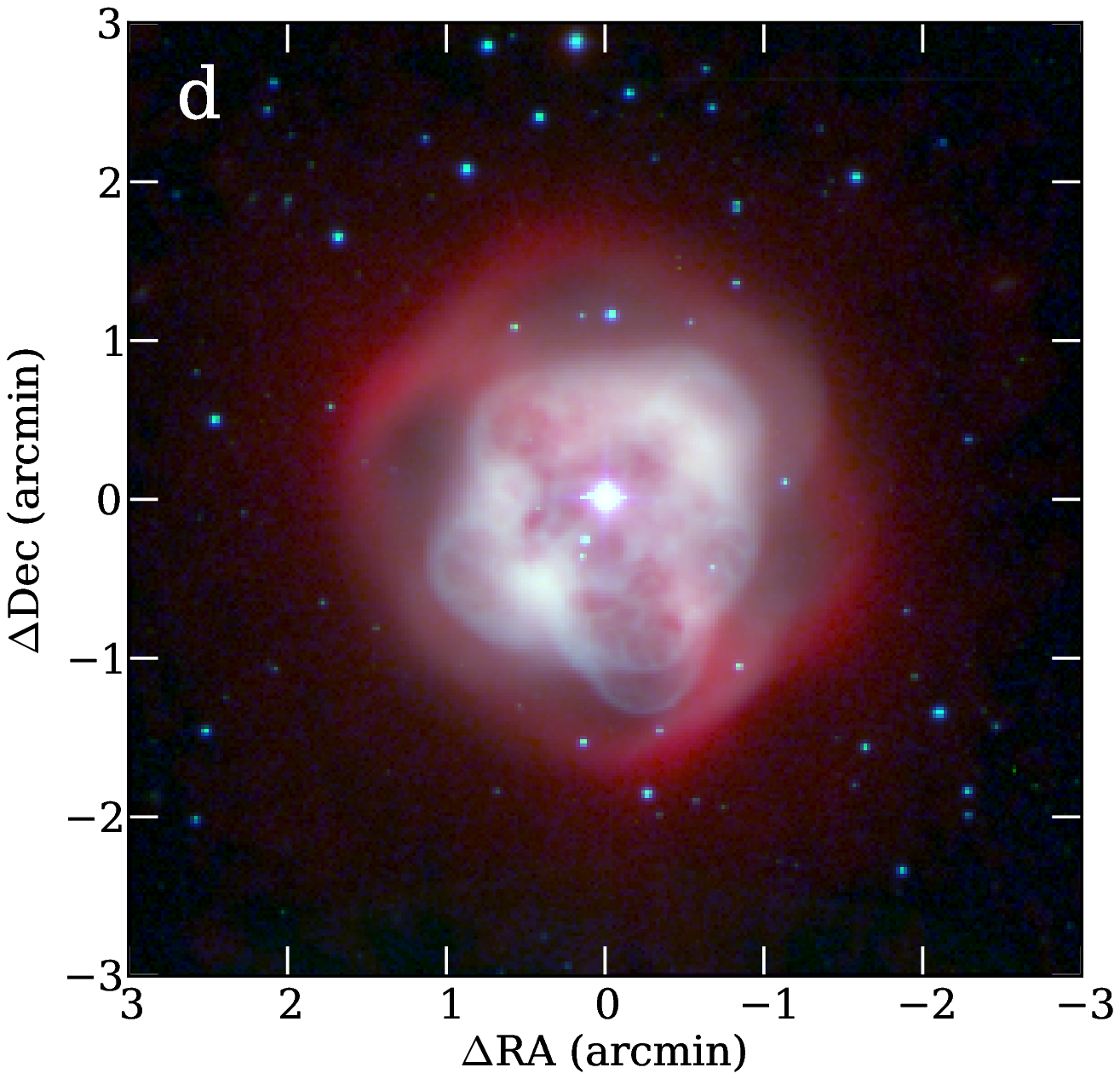} \\
\end{tabular}
\end{center}
\caption{NGC~1514 as seen in various combinations of WISE and visible
  wavelength filters. The axes indicate arcminute offsets from the
  nominal central source position of $\alpha = 04^{\mathrm h}$ 09$^{\mathrm
    m}$ 16\fs990, $\delta =$~+30\degr{} 46\arcmin{} 33\farcs44 (J2000). a) WISE
  3-color image, blue = W2 [4.6 \microns], green = W3 [12 \microns]
  and red = \decon{} [22 \microns], normalized to bring out the
  different infrared features. b) Visible 3-color image derived from
  POSS B, R, and I images, blue = B, green = B+R, red = R+I; rebinned
  to WISE's pixel scale and normalized so that the field stars are
  white. c) Visible wavelength combination of blue = POSS B, green =
  [\ion{O}{3}], and red = H$\alpha$, normalized so that the nebulosity
  at (0.2, -0.6) is white. d) Combination of blue = POSS B, green =
  H$\alpha$, and red = W3, also normalized so that the nebulosity at
  (0.2, -0.6) is white. From c) and d), we conclude that the inner
  shell is remarkably ``gray'' across all wavelengths, in marked
  contrast to most other PNe.
\label{fig:wise}}
\end{figure*}

Inspection of even the raw, unprocessed WISE images immediately
revealed an unexpected structure in NGC~1514. A pair of bright
axisymmetric rings that surround the visible nebula are plainly
apparent (coadded data are shown in Figure~\ref{fig:wise}a); in fact,
they dominate the emission at W4, giving the red appearance to the
rings in the figure. Such a structure is not suggested in any of the
visible wavelength images available to us; \eg\ the DPOSS data
presented in Figure~\ref{fig:wise}, the H$\alpha$ image in Figure
\ref{fig:guide}, the published figures of \citet{balick1987},
\citet{hajian1997}, and \citet[and private
  communication]{corradi2003}, or the deep, very narrow-band image by
D.~Goldman\footnote{\url{http://dg-imaging.astrodon.com/gallery/display.cfm?\-imgID=205}}.

Remarkably, the inner shell has almost exactly the same appearance in
the mid-infrared as in the visible when the differing spatial
resolutions are taken into account. Panels c and d of Figure
\ref{fig:wise} show that the inner shell is ``gray'' across all
wavelengths (with the exception of the rings). The spatial extent, the
positioning of the bubbles, the relative brightnesses, etc.\ are all
identical at the $\sim$ 10--20\% level over many different
wavelengths. This is in marked contrast to other PNe such as those
illustrated in \citet{balick2002} and \citet{hora2004}, though the
Group 1 PNe of \citet{chu2009} have 24 \micron\ morphologies that
compare favorably to the optical red or H$\alpha$ images. Images of these objects at
other wavelengths are needed to determine if they are similar
to NGC~1514 in this regard.

\subsection{Photometry}

The spatial resolution of WISE does not allow much new insight
directly into the nature of the central source of NGC~1514; the
6\arcsec{} resolution has no hope of resolving the central
binary. If the sdO+A0III (or horizontal branch A0) binary model is
correct, then at best we would expect the infrared fluxes to
approximately follow the Rayleigh-Jeans law for a $\sim$ 10,000 K
blackbody. To verify this, we present infrared fluxes from the 2MASS
and WISE surveys in Table \ref{tab:mags}. These values are obtained by
profile fit photometry as returned by the 2MASS and WISE data
pipelines. Encouragingly, \citet{taranova2007} got very similar
magnitudes in the near-infrared using an aperture photometer: J =
8.238, H = 8.108, K = 8.036, and L = 7.927.

\begin{table*}
\begin{center}
\caption{Flux of the Central Source of NGC~1514\label{tab:mags}}
\medskip{}

\begin{tabular}{c c c c r@{.}l r@{.}l c c}
 &  &  &  & \multicolumn{4}{c}{Observed} & \multicolumn{2}{c}{Dereddened}\\
Band & $\lambda$ & 0 Mag & Extinc.& \multicolumn{2}{c}{Magn.}  & \multicolumn{2}{c}{Flux} & Magn. & Flux  \\
 &           & Flux   & Corr.   &\multicolumn{2}{c}{}       & \multicolumn{2}{c}{}     &      &       \\
 &  (\microns) & (Jy) & (mags)  & \multicolumn{2}{c}{(mags)}& \multicolumn{2}{c}{(Jy)} & (mags)& (Jy) \\
\hline 
J  & 1.235  & 1594  & 0.442 & 8&190 & 0&844 & 7.748 & 1.268 \\
H  & 1.662  & 1024  & 0.275 & 8&098 & 0&590 & 7.823 & 0.761 \\
K  & 2.159  & 666.7 & 0.133 & 8&002 & 0&420 & 7.869 & 0.475 \\
W1 & 3.353  & 306.7 & 0.058 & 7&918 & 0&209 & 7.860 & 0.220 \\
W2 & 4.603  & 170.7 & 0.034 & 7&907 & 0&117 & 7.873 & 0.121 \\
W3 & 11.561 & 29.04 & 0.048 & 7&576 & 0&027 & 7.528 & 0.028 \\
W4 & 22.088 & 8.239 & 0.026 & $>$ 7&5 & $<$ 0&008 & ---   & ---   \\
\end{tabular}
\end{center}
\end{table*}

The visual extinction to NGC~1514 has been estimated by various
authors to be 1.5 to 2.5 magnitudes, based primarily on the $E(B-V)$
color \citep[\eg][]{greenstein1972} or from the H$\alpha$/H$\beta$ or
radio/H$\beta$ ratios \citep[\eg][]{cahn1992}. We use the new WISE
fluxes to provide a much longer baseline in an effort to better
estimate the correct visual magnitude of NGC~1514 photometrically. We
begin by constructing a model CSPN spectrum using visible-wavelength
spectral templates for an O5V star and an A0III star \citep[from
  STELIB,][]{leborgne2003}, and a 9800 K blackbody to represent the
infrared. The precise spectral types are relatively unimportant; we
are primarily interested in correctly estimating the overall shape of
the SED, not in the particulars of any spectral features. Using the
interstellar extinction curves of \citet{drainelee}, we redden the
model until an acceptable match is achieved. (Extinction curves of
other authors were checked, but the differences are not large enough
to affect our conclusions, and \citeauthor{drainelee} are the only set
comprehensive enough to cover ultraviolet through infrared
wavelengths.)

\begin{figure}[!ht]
\begin{center}
\includegraphics[width=0.48\textwidth]{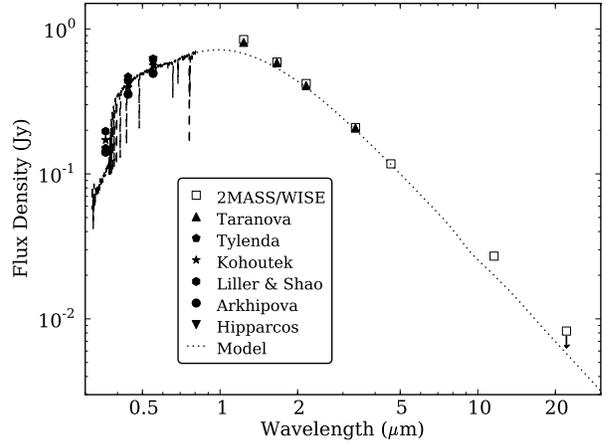}
\end{center}
\caption{The SED of the central source. The plot shows a composite
  spectrum of an O5V template (HD46223), an A0III template (HD123299),
  and a 9800 K blackbody, reddened with $A_V=1.6$ (dashed
  line). This is overlayed on the WISE and 2MASS data (open squares), and a
  selection of UBV and near-IR photometry (\citealt*{arkhipova1969},
  \citealt*{kohoutek1967}, \citealt*{liller1968},
  \citealt*{taranova2007}, \citealt*{tylenda1991}). Except for the $J$
  flux, the SED of the CSPN is well represented by this composite
  spectrum and gives us confidence that the intrinsic visual magnitude
  of the central source is 7.86. \label{fig:sed}}
\end{figure}

The best model fit is shown in Figure~\ref{fig:sed}; an extinction
value of $A_{V}=1.6$ is found. The fit is quite good except for the J
band near-infrared point and the W3 mid-infrared point. Given the good
agreement between the 2MASS profile fit value and the aperture-derived
value of \citet{taranova2007}, the ``excess'' at J is real, at least
in the sense that it is not a measurement error and it cannot be fit
with our simplistic photospheric model. \citet{whitelock1985} forecast
the great strength of the \ion{He}{1}\ triplet at 1.083 \microns\ in
many PNe, while Pa$\beta$ can also be very strong
\citep[\eg][]{hora1999}, so it is possible that some nebular emission
is being picked up where it overlaps the CSPN. (Though to be fair, the
K band will pick up Br$\gamma$ and may also be affected.) The small
excess at 12 \microns{} is also real, and is definitely due to the
nebular line emission (see Section \ref{sec:spitzer}).

Using this value of $A_V$, we deredden the magnitudes for the CSPN,
and these are also reported in Table \ref{tab:mags}. If we assume that
the near-infrared SED is indeed that of an A0 star where all the color
terms are 0.0, we should find the same magnitude value at all
wavelengths. Taking the mean of the magnitudes from H through W2, the
dispersion is reduced from 7.98 $\pm$ 0.08 for the uncorrected case
(where there is a obvious slope to the data) to 7.86 $\pm$ 0.02 after
dereddening. The literature shows a fair spread in the reported $B$
and $V$ band values (though $B-V$ is pretty consistently between 0.50 and 0.55),
but an average of a half dozen reported values shows that
$V=9.53\pm0.09$. Applying our value of $A_{V}$ to this mean corrects
the $V$ magnitude to 7.93, only 7\% off the 7.86 found from the
infrared magnitudes. In subsequent discussion, we assume that the best
estimate of the intrinsic V magnitude of the CSPN is $m_V=7.86$.

\subsection{Surface brightness}
\label{sec:redness}
\begin{figure}[t]
\begin{center}
\includegraphics[width=1\columnwidth]{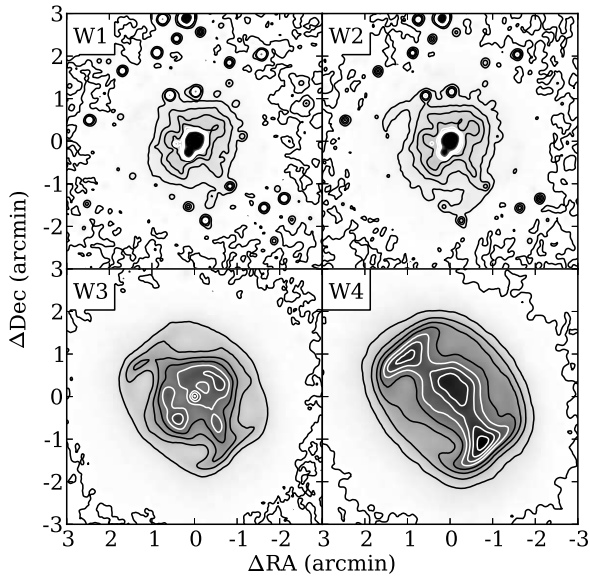} 
\end{center}
\caption{Surface brightness contour plots of NGC~1514 at the four WISE
  wavelengths. Contours for W1 and W2 range from 0 to 0.8 (black) and
  1.0 to 1.8 (white) MJy~sr$^{-1}$ in steps of 0.2; for W3: 0 to 8
  (black) and 10 to 18 (white) MJy~sr$^{-1}$ in steps of 2; and for
  W4: 0 to 16 and 20 to 32 MJy~sr$^{-1}$ in steps of
  4.\label{fig:all}}
\end{figure}

Contour maps of NGC~1514 at the four WISE wavelengths are presented in
Figure~\ref{fig:all}. The nebular emission is clearly rising rapidly
at longer wavelengths. The surface brightness of the nebula in the
region of the bubbles roughly follows a power law $F_\nu \propto
\nu^{-2.0}$ for the four WISE bands. (This holds only for the WISE
bands; the inner shell is brighter in the 2MASS bands than this power
law would predict. Our assumption is that shorter wavelengths are
dominated by line emission and perhaps scattered starlight.) It is
also the case that the emission from the rings is rising even more
rapidly (is redder) than that in the inner core, following a power law
of $F_\nu \propto \nu^{-2.8}$.

Therefore, the nature of the nebular emission at mid-infrared
wavelengths appears to be primarily thermal, even in regions where the
line strength is relatively high. Though the emission from the
brightest mid-infrared emission line, that of the [\ion{O}{4}] line at
25.89 \microns, is reasonably strong (see Section \ref{sec:spitzer}),
it lies near the long wavelength cutoff of the W4 passband where the
transmission is low, and thus cannot account for a substantial
fraction of the observed flux.

\begin{figure}[!ht]
\begin{center}
\includegraphics[width=\columnwidth]{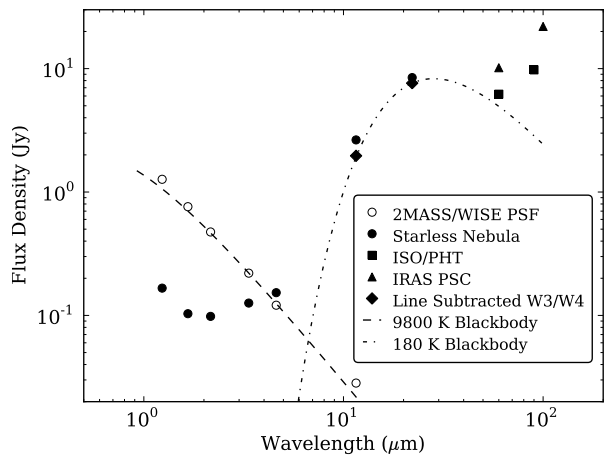}
\end{center}
\caption{A plot of the integrated nebula SED using the dereddened
  2MASS, WISE, ISO/PHT, and IRAS fluxes. The central source flux
  (found by PSF fitting) is shown for comparison and is represented by
  the open circles, the integrated nebular flux (see text) is given by
  the filled circles, and the ISO and IRAS point source fluxes are
  indicated by the filled squares and triangles. An estimate of the
  contribution of the atomic lines was subtracted from the W3 and W4
  fluxes, and this net thermal emission is indicated by the filled
  diamonds. A 9800~K blackbody curve is overlayed on the CSPN fluxes
  while a 180~K blackbody is drawn through the WISE line-corrected W3
  and W4 fluxes to suggest that the character of the mid-infrared
  nebular emission is thermal in nature.
\label{fig:intneb}}
\end{figure}

Figure~\ref{fig:intneb} shows a plot of the total integrated flux of
the nebula (all stars, including the CSPN, have been removed by profile
fitting) from 1 to 100 \microns, measured using a 5~arcminute diameter
aperture for the 2MASS, WISE, and NVSS data, 3.5~arcminutes for the
ISO/PHT data (the maximum extent available in the released data), and
the IRAS point source catalog's values where we assume all of the flux
will be captured in a single beam. These values are also listed in
Table \ref{tab:extra}.

\begin{table*}
\begin{center}
\caption{Infrared Integrated Nebula Fluxes\label{tab:extra}}
\begin{tabular}{c c c c}
Source/& Wavelength & Observed & Dereddened\\ 
Band   &            & Flux Density& Flux Density\\
       & (\microns) & (Jy) & (Jy) \\ 
\hline 
J       & 1.24   & 0.111 & 0.166\\
H       & 1.66   & 0.080 & 0.103\\
K       & 2.16   & 0.087 & 0.098\\
W1      & 3.35   & 0.119 & 0.126\\
W2      & 4.60   & 0.148 & 0.153\\
W3      & 11.56  & 2.529 & 2.644\\
W4      & 22.09  & 8.264 & 8.461\\
ISO/PHT & 60     &  6.205& ---\\
ISO/PHT & 90     &  9.775& ---\\
IRAS    & 60     &  10.17& ---\\
IRAS    & 100    &  21.97& ---\\
\end{tabular}
\end{center}
\end{table*}

At wavelengths shorter than 4.6 $\microns$, the CSPN has a higher flux
than the whole of the nebula. But at wavelengths longer than 4.6
$\microns$, the nebula's flux density rises dramatically. A 180 K
blackbody curve has been drawn into the figure to guide the eye, but
it is clear that multiple temperature components are required to
completely fit the data. A more detailed thermal model is not
warranted given the few data points, but it is the case that the bulk
of the emission comes from material that is quite cool, $T \lesssim$
200 K, and some perhaps as low as 30 K to account for the far-infrared
fluxes.

\subsection{Background}

\begin{figure}[!ht]
\begin{center}
\includegraphics[width=1\columnwidth]{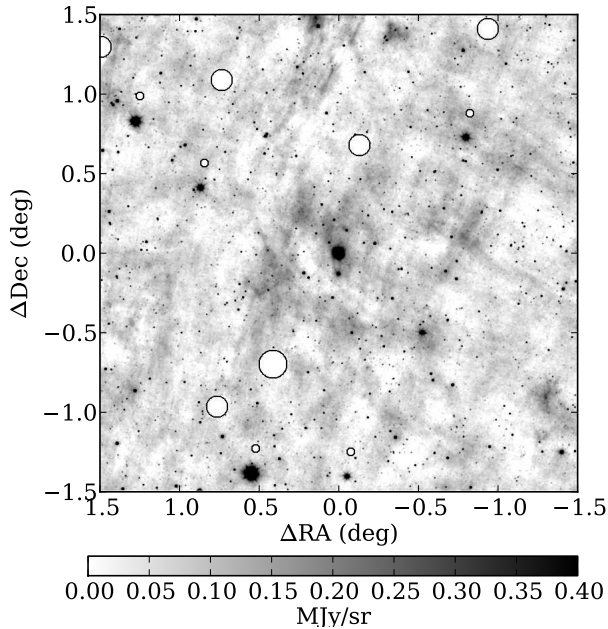}
\end{center}
\caption{A 3$\times$3 degree field in the vicinity of NGC~1514 shows
  the striated background visible in the W3 band. The ridge upon which
  the PN is projected has a brightness of about 0.34 MJy~sr$^{-1}$
  above the darkest area of the region (which may not be zero flux).
  The PN itself is at the center. The white circles mask areas with
  known image defects: detector latents and optical reflections, both
  caused by very bright stars. The equivalent W4 image, except for the
  factor-of-two lower spatial resolution, has all the same structures
  that are visible here.\label{fig:cirrus}}

\end{figure}

There is a moderately bright, dusty background present in the region
near NGC~1514; \citet{schlegel1998} (as implemented by the NED
Extinction Calculator) predict an $A_V$ of 2.252 magnitudes in this
direction ($l$~=~165.53, $b$~=~-15.29). It appears that the PN is
superimposed on top of one of the brighter ridges of this background;
we estimate a 12 \micron\ brightness of 0.34 MJy~sr$^{-1}$ at the PN
above the darkest areas of the image. (The zero point is not yet
calibrated, so we cannot know the absolute brightness.) A 3\arcdeg$\times$3\arcdeg\ W3 field is shown in Figure~\ref{fig:cirrus}. The 
equivalent W4 field shows all the same structures visible at W3, while a
comparison field taken 45\arcdeg\ farther off the galactic plane
($l$~=~165.53, $b$~=~-60.29), where the predicted $A_V$ is only 0.076,
shows no such background structure. We are therefore confident that
this background structure is real.

The extended emission outside a 5 arcminute diameter circle around NGC
1514 is dominated by this striated background, and we see no obvious
correlation between these structures and the nebula, either in
position or symmetry. The PN is simply superimposed onto one of the
denser portions of this region. This conclusion calls into question
some of the results found by \citet{weinberger1999} and
\citet{aryal2010} where they reported structures with degree angular
scales that were linked with NGC~1514 in the IRAS data at 12
\microns{}. The emission \citeauthor{aryal2010}\ see in their Figure~1c 
matches that of our Figure~\ref{fig:cirrus}, but the much poorer
spatial resolution of IRAS has smeared the background variations into
the ring-like structure those authors report. Indeed, if we bin the
WISE data down to two arcminute pixels, the appearance is similar
enough to suppose that there is a link. However, the factor-of-60
higher resolution WISE data show that the background nebulosity is not
correlated with the PN in any simple way.

\subsection{Spitzer spectra}
\label{sec:spitzer}

\begin{sloppypar}
The Spitzer Space Telescope observed NGC~1514 only once as part of a
wavelength calibration program for the IRS instrument. Accordingly,
the exposures are very short, ranging from 15 seconds in the
low-resolution data to 60 seconds in the high resolution data, so that
the overall signal-to-noise ratio is small. In addition, the pointings
between the four different spectrometer modules do not overlap, so
that only on the central source do they all sample the same region.
These data are publicly available and were accessed through the
Spitzer Data Archive.
\end{sloppypar}

\begin{figure}[!ht]
\begin{center}
\includegraphics[clip,width=0.49\textwidth]{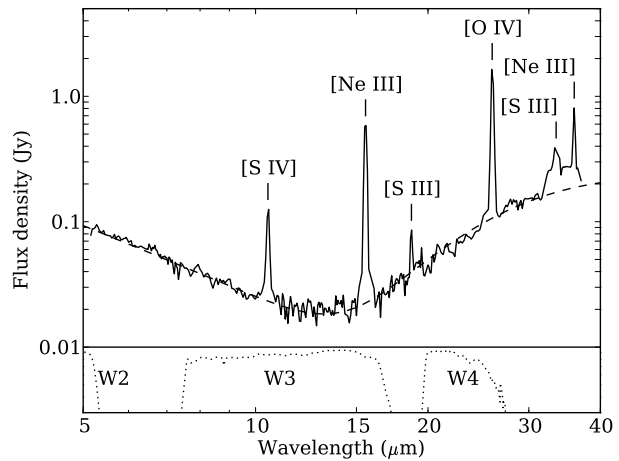} 
\caption{Spitzer low resolution spectrum at the position of the CSPN.
  Both stellar and nebular contributions are included in the flux. The
  dashed line represents a 9800 K blackbody photosphere plus a 100 K
  dust continuum. The WISE relative spectral response curves (W1 is
  off scale to the left) are plotted below the spectrum for reference;
  the scale is logarithmic from 1\% to 100\%. Only the six labeled
  forbidden atomic lines are detected; the 10.51 \micron{}
  [\ion{S}{4}] and 15.56 \micron{} [\ion{Ne}{3}] will contribute to
  the W3 flux, but the 25.89 \micron{} [\ion{O}{4}] line will
  contribute little to the W4 flux since the response is falling off
  rapidly. No significant PAH emission is seen.\label{fig:spitzerlo}}
\end{center}
\end{figure}

Low resolution spectra were extracted for the central position by
using the most central of the pointings while using the most
off-center pointings to subtract the background spectrum. This
spectrum is presented in Figure~\ref{fig:spitzerlo}. Several facts are
immediately apparent: 1) the short wavelength end of the IRS spectrum
is dominated by emission from the 9800 K photosphere; 2) the long
wavelength end is dominated by 100~K thermal emission from
circumstellar material; 3) forbidden atomic lines from the nebula are
quite strong even in just the line of sight to the CSPN; and 4) there
is a complete lack of the PAH emission bands. Although PAH emission is
usually dominant in nebulae, many PNe are known to have no PAH band
emission and are instead dominated by H$_2$ line emission; NGC~1514
appears to be a member of this class. Since the WISE W3 band
encompasses all the PAH emission bands from 7.6 to 13.5~\microns, the
Spitzer data suggests that all the structure seen in W3 arises from
the same material that produces the optical nebula, along with thermal
emission from dust that is distributed throughout.

\begin{figure}[!ht]
\begin{center}
\includegraphics[clip,width=0.49\textwidth]{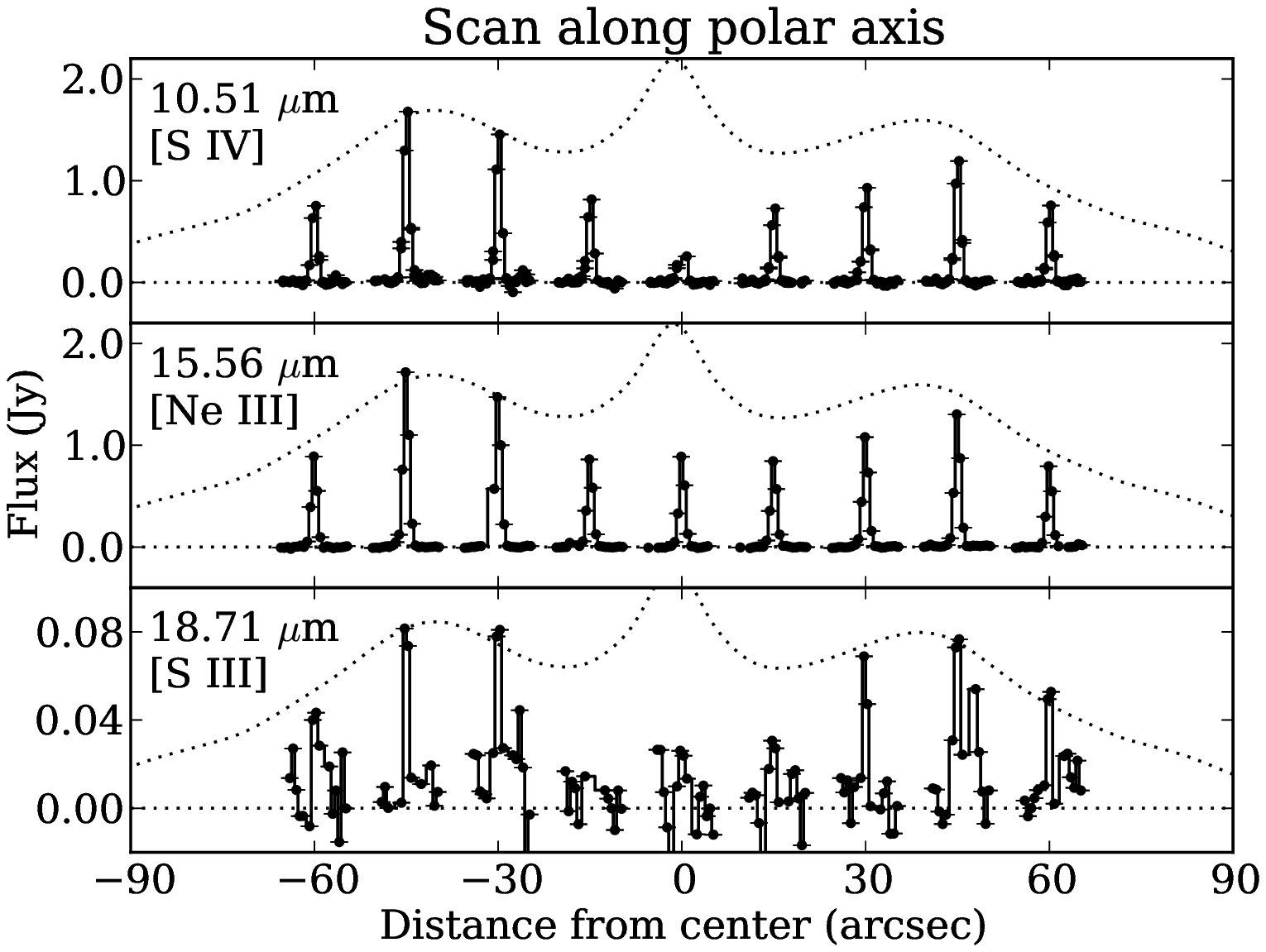} 
\smallskip

\includegraphics[clip,width=0.49\textwidth]{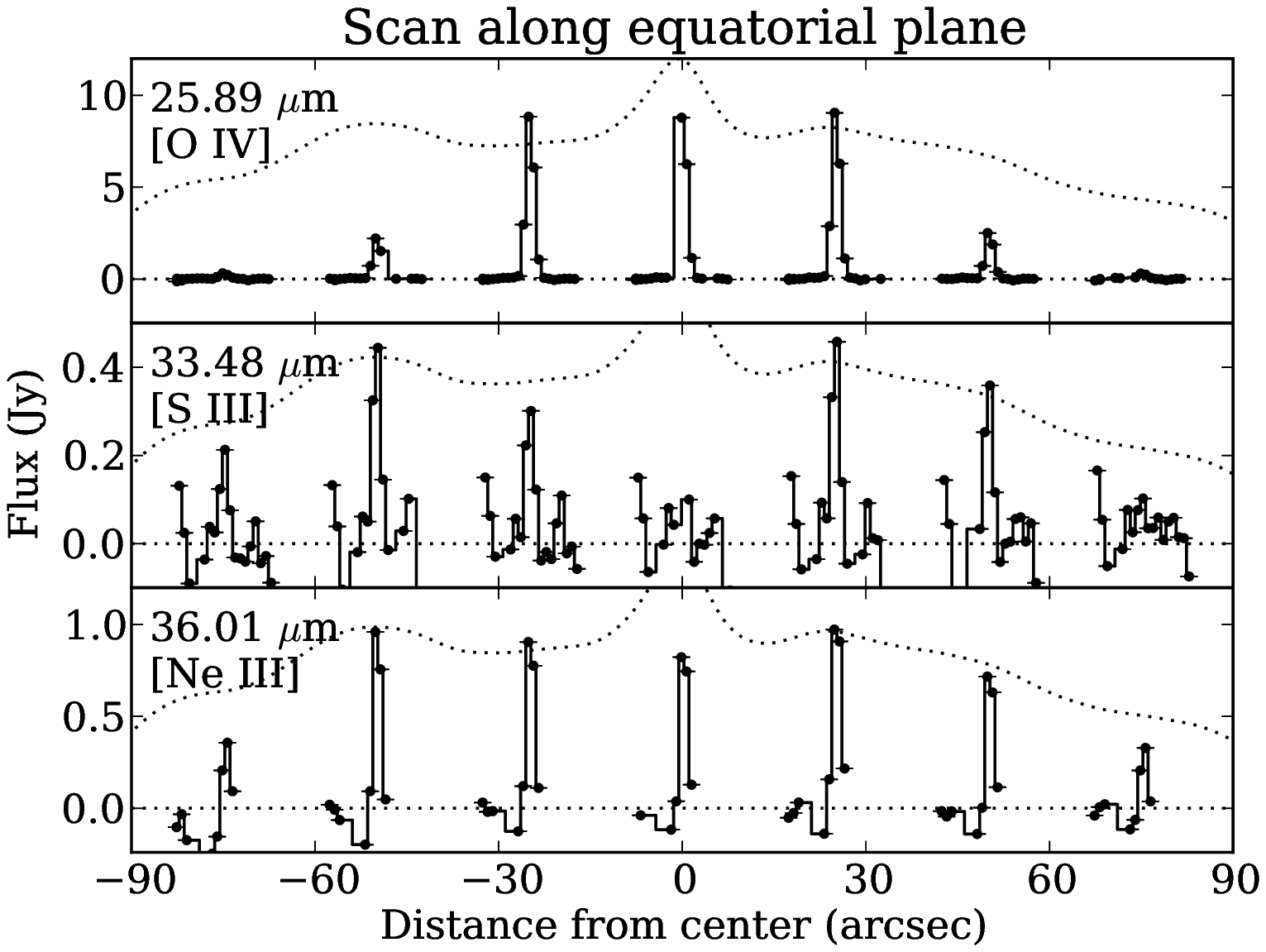} 
\end{center}
\caption{Spitzer high resolution spectra. Only these six lines are
  detected in the very short exposures. The spatial variations in
  intensity of five of the six lines match the intensity profile of
  the broadband W3 filter except at the center where the CSPN
  dominates the broadband profile (an arbitrarily-scaled version is
  plotted as the dotted curves). Only in the [\ion{O}{4}] line does
  the distribution vary from the profile; it is concentrated in the
  central 1-arcminute-diameter region.\label{fig:spect}}
\end{figure}

Only six emission lines are unambiguously detected at either spectral
resolution: the 25.89~\micron{} [\ion{O}{4}] line is quite strong; the
10.51~\micron{} [\ion{S}{4}], 15.56~\micron{} [\ion{Ne}{3}], and 36.01~\micron{} [\ion{Ne}{3}] lines are fainter, but still well detected;
and the 18.71~\micron{} and 33.48~\micron{} [\ion{S}{3}] lines are
very faint, but definitely present. The presence of these lines, in
particular that of [\ion{O}{4}], indicates a high excitation nebula
(central source of order 100,000 K), though not as high as some
because of the lack of the [\ion{Ne}{5}] line at 24.32~\microns.

Nine spatial positions along the PN's polar axis were sampled at
high spectral resolution with the Short-High module, and seven
positions were obtained along the equatorial plane with the Long-High
module (Figure~\ref{fig:spect}). We note that, with the exception of
the [\ion{O}{4}] line, all the lines vary in intensity as a function
of position in the same manner as the broadband W3 flux; the W3
intensity profile matches the envelope of the line peaks surprisingly
well. This implies that at least the sulphur (doubly and triply
ionized) and neon are well mixed with the dust throughout the inner
shell. The oxygen, on the other hand, (at least the triply ionized
species) is concentrated in the inner 30\arcsec\ radius of the nebula,
and is almost entirely absent outside that.

The outer two positions taken with Long-High module (75 arcsec from
center) are the only two positions that would be dominated by flux
from the rings. In both positions, the lines strengths (25.89, 33.48,
and 36.01 \microns) are much less than for the more central positions.
Given that the nearest WISE band, W4, shows that the rings are
brighter than the inner shell at long wavelengths, this suggests that
lines are responsible for little of the ring emission; the rings must
be almost entirely thermal emission.

Because no pair of emission lines from the same species was observed
at the same spatial position, we can only make crude estimates of line
ratios. One of the better indicators of electron density is the
[\ion{S}{3}] 18.7/33.5 \micron\ line ratio. If we make the assumption that
the underlying nebular brightness is roughly constant 50\arcsec\ from
the CSPN, then the ratio is a very low 0.33 ($2.2\times 10^{-17}$
W~m$^{-2}$ vs. $6.6\times 10^{-17}$ W~m$^{-2}$). Even if our uniformity
assumption is off by more than a factor of two either direction, the
inferred electron density is still well under $10^3$ cm$^{-3}$. This
agrees with the estimate by \citet{kohoutek1967} who derived a value
of 290 cm$^{-3}$ for the inner shell and 140 cm$^{-3}$ for the outer
shell based on H$\beta$ line strengths.

\section{Discussion}

Given that the central source of NGC~1514 is a binary system, the
presence of these rings, while perhaps surprising in their form, is not
entirely unexpected. Binarity and bipolar structures in PNe go hand-in-hand
\citep[\eg\ the review by][and references therein]{demarco2009}. The
new infrared data, and particularly the discovery of the rings, allow us
to refine our understanding of NGC~1514.

\subsection{Distance}

The distance to NGC~1514 is quite uncertain. From the Hipparcos
catalog of parallaxes, one can derive a distance of $185\pm58$ pc,
while the statistical methods summarized by \citet{zhang1995} can
yield distances up to 1300 pc. Our WISE results do not shed a great
deal of new light on the distance, but we can use our more confident
estimate of $m_V=7.86$ to help set plausible limits on the range of
possibilities.

In deriving the distance modulus to the nebula, the absolute magnitude
of the CSPN has a larger uncertainty by far than the dereddened
apparent visual magnitude. If the brighter component of the CSPN is
taken to be an A0III giant as determined by \citet{kohoutek1967} with
an absolute magnitude of $\sim -1$ \citep{jaschek1998}, then the
inferred distance is $\sim 600$~pc. If, instead, the CSPN is a
horizontal branch A star as found by \citet{greenstein1972} with an
absolute magnitude $\sim +1$ \citep{deboer1997}, then the distance is
$\sim 240$~pc.

It seems likely, therefore, that distances $> 600$~pc such as those
found by the H$\beta$-to-radio ratio \citep{cahn1992} or
diameter-to-radio flux relationships \citep{daub1982} are
overestimates. The Hipparcos distance (at $+1\sigma$) is broadly
consistent with our lower luminosity estimate, thus we consider a
distance of 200--300~pc quite plausible.

\subsection{Ring Properties}

\begin{figure}[!ht]
\begin{center}
\includegraphics[width=1\columnwidth]{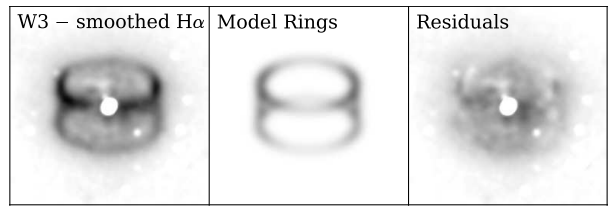}
\end{center}
\caption{The residuals of the ring model fit for the W3 image. A
  smoothed version of the H$\alpha$ image has been subtracted to
  remove the inner shell from the W3 image (left panel); this image
  was then rotated clockwise 131\arcdeg\ so the polar axis is
  vertical. The ring model (center) was then subtracted to reveal the
  residuals (right). While not a perfect fit due to the
  over-simplicity of the model, the general orientation and aspect
  ratios seem to be correct.\label{fig:ringfit}}
\end{figure}

For purposes of extracting geometric information, the rings can be
modeled as two parallel, unresolved rings (Figure
\ref{fig:ringfit}) with a deprojected
separation of 41\arcsec{} (0.05 pc, if we assume $d=250$~pc), the southeast of which is 173\arcsec{} in
diameter while the northwest is 177\arcsec{} in diameter (roughly 0.2 pc). The rings are tilted 59\arcdeg{} from pole-on and
rotated to a position angle of 131\arcdeg. A formal least squares fit
using a downhill simplex method was attempted, but the irregularities
in the ring brightness and the complexity of the interior shell
morphology prevented a believable solution from being found. The
values quoted are estimated from ``by eye'' fits, but numerous trials
showed noticeable errors in the diameter and separations at the
$\pm1$\arcsec{} level, in PA at $\pm0.5$\arcdeg, and in tilt at
$\pm2$\arcdeg. Higher spatial resolution imaging than can be provided
by WISE will be required to refine these estimates.

By subtracting the ring models from the integrated flux measurements,
we estimate the percentage contribution of the rings to the total
nebular flux. The rings contribute $\lesssim$~10\% at W1 and W2,
$\sim$~15\% at W3, and $\gtrsim$~30\% at W4. The increasing fraction
is consistent with the redder color of the rings noted in Section
\ref{sec:redness} and suggests a ring color temperature of
$\sim 160$~K.

\begin{figure*}
\begin{center}
\includegraphics[width=1\textwidth]{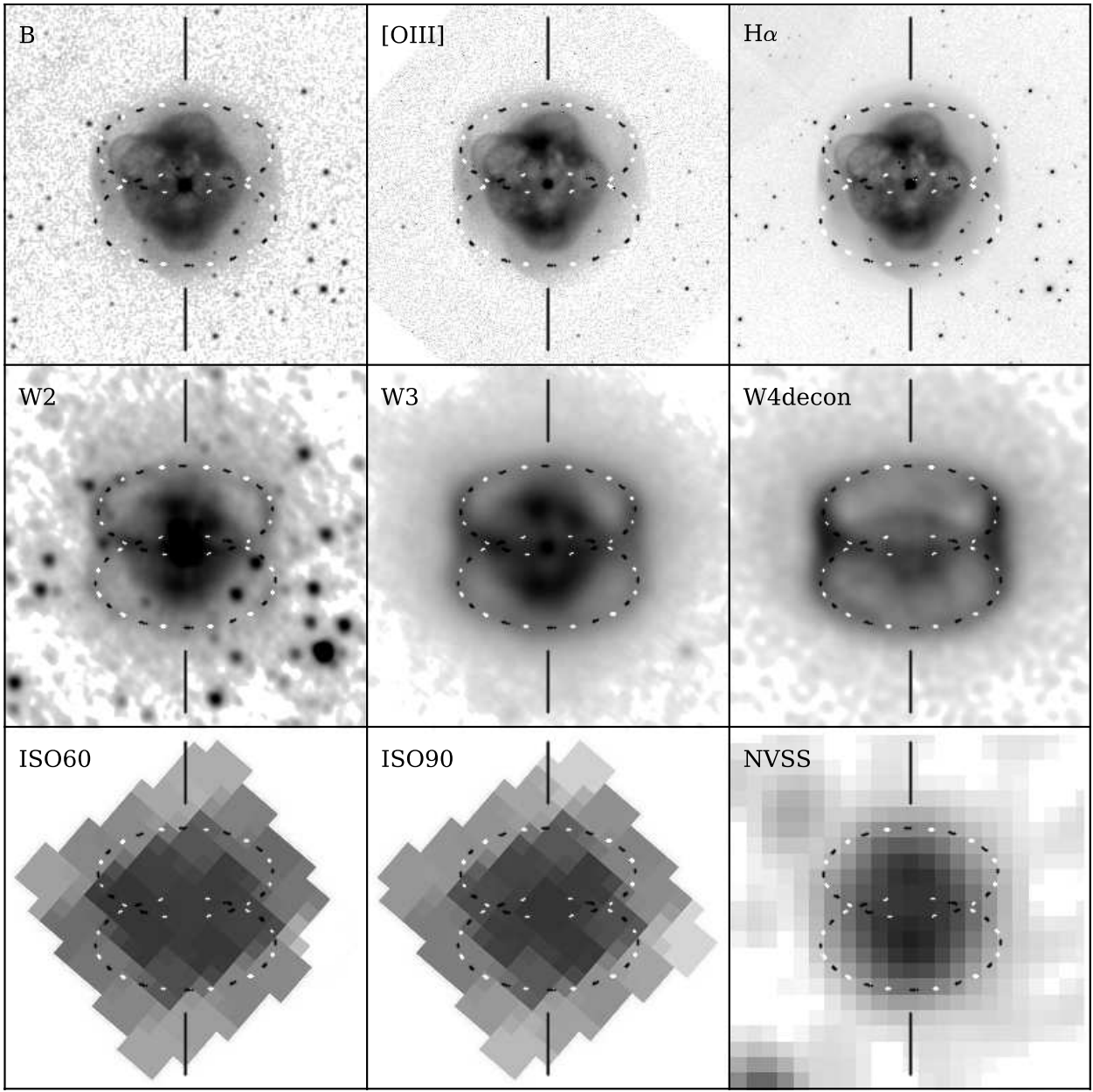}
\end{center}
\caption{A comparison of a double ring model (drawn as the alternating
  black and white dotted curves) to the POSS B, \ion{O}{3}, and
  H$\alpha$ data (top row, left to right), the W2, W3, and W4 images
  (middle row), and the ISO 60 \micron, ISO 90 \micron, and 1.4 GHz
  NVSS maps (bottom row). It is clear that the rings are contained
  entirely within the outer shell.\label{fig:rings}}
\end{figure*}

This ring model fit can then be overlayed on all the images in order
to see how the various features visible at differing wavelengths
compare (Figure \ref{fig:rings}). Starting with the middle row, the
fit falls well on the rings seen at all four WISE wavelengths (W1 is
very similar to W2 and has been omitted). At visible wavelengths, the
ring fit falls entirely within the faint outer shell, but no evidence
can be seen for the rings themselves at these wavelengths. In
the far-infrared, the nebula is not well resolved, but it appears that
it is elongated roughly along the equatorial plane of the model,
suggesting that there is an extensive, dusty disk lying in the plane.
This disk cannot be very dense given that the extinction to the
central source is not large, and there are no color gradients across
the nebula.

At 1.4 GHz in the NVSS survey, the nebula is resolved into two spots
corresponding to the visible bubbles. \citet{pazderska2009} report a
total flux of 60~mJy at 30~GHz for NGC~1514. They note that this PN
shows no sign of a high-frequency excess such as that attributed to
emission from spinning dust \citep{casassus2007}. The radio emission
from NGC~1514 is therefore purely free-free; this again suggests that
the bubbles contain a significant quantity of ionized gas while the
rings do not.

\subsection{A Physical Model, Refined}

\citet{muthu2003} present a physical model of NGC~1514 in their
Section 4 that agrees very well with our observations. The overall
ellipsoidal shape of the PN is still apparent in the infrared. The
presence of the rings confirms the axisymmetric structure they assume.
We refine the estimate of the position angle of the polar axis from
135\arcdeg\ to 131\arcdeg, and we establish a tilt angle of
31\arcdeg\ out of the plane of the sky, agreeing with them that the
southeast part of the nebula is the side that opens toward us. While
they could not determine the relative angles between the bubble axis
and the shell axis, we can confirm that they are coincident (see, for
example, Figure~\ref{fig:rings}). This general configuration now seems
quite secure.

The formation of the rings themselves is still uncertain. At the
simplest level, the rings are part of the dust that is associated with
the photon-dominated region (PDR) that is, in turn, wrapped around the
ionized zone. But, lacking specific kinematic information on the rings
(recall that \citealt{muthu2003} concentrated on the inner shell, and
where their sample points might have contained the rings, the rings
are not detected), we can only compare NGC~1514 to other objects that
have similar morphologies. Two such objects are Hen~2-104 and MyCn~18,
although in both of these cases the structures are seen at visible
wavelengths.

The core of Hen~2-104 bears the most striking resemblance to NGC~1514,
though Hen~2-104 itself is a symbiotic star rather than a PN and it
shows multiple sets of rings. \citet{corradi2001}, and updates by
\citet{santander2008}, develop spatiokinematic models to explain the
rings and accompanying structures. They invoke the model proposed by
\citet{soker2000} that uses interacting winds from the central binary
source to explain the multiple rings and overall collimated structure.
Such a model could partially explain NGC~1514 if we assume that the
rings we observe represent a particularly large mass loss event and
are thus merely the brightest ones in the system, and that other rings
and a more extensive structure may be detectable if greater
sensitivity were available to us (and it were not washed out by the
dusty background emission). However, the focusing mechanism of the
winds cannot be so efficient as to prevent the formation of the
relatively uniform outer shell seen at visible wavelengths.

The formation of MyCn~18 may follow similar lines. It has a very well
defined hourglass shape with numerous pairs of parallel rings, though
the overall structure is much more focused than either Hen~2-104 or
NGC~1514. \citet{sahai1999} propose a mechanism related to that of
\citet{soker2000}, except that the rings are formed by a more
collimated outflow jet interacting with a round circumstellar
envelope. This could again fit NGC~1514 if we assume that there was
one event that was far stronger than any other, and any other events
are simply too faint for WISE to detect.

This still leaves the question of why the rings are visible only in
the infrared. Our interpretation is that, based in the spectral index
of the ring emission discussed earlier, the rings are still present
but are just too faint in comparison with the more uniformly
distributed line emission in the outer shell. The rings are simply
washed out at visible wavelengths.

\section{Conclusion}

The discovery of axisymmetric rings in NGC~1514 completes our
perception of it as a planetary nebula formed from an aging binary
star system. No longer can NGC~1514 be considered a simple planetary
nebula ``of a circular form'' as Herschel perceived it. While the
morphology is very complex, with numerous bubbles contained within the
inner shell and rings contained within the outer shell, it joins the
family of hourglass-shaped nebulae.

It is likely that more such unexpected structures will be found in PNe
as the WISE survey data becomes widely available. As new
hourglass objects are found, the formation models can be put to more
rigorous tests, and we will gain a deeper understanding of how these
mysterious and beautiful objects have come to exist.

\acknowledgements

We thank our anonymous referee for the very thorough review of this
manuscript and the suggestions which improved our discussion. We are
grateful to Romano Corradi for his generous provision of the FITS
files for the \ion{O}{3} and H$\alpha$ images of NGC~1514, and to Roc
Cutri and Frank Masci for explanations of the WISE processing
pipeline. This research was carried out at the Jet Propulsion
Laboratory, California Institute of Technology, under a contract with
the National Aeronautics and Space Administration. M.C.\ thanks NASA
for supporting his participation in this work through UCLA Sub-Award
1000-S-MA756 with a UCLA FAU 26311 to MIRA. Finally, we wish to
acknowledge the many scientists and staff of all the WISE partners who
spent many years making WISE successful.

This research has made use of the SIMBAD database, operated at CDS,
Strasbourg, France, and the NASA/IPAC Infrared Science Archive, which
is operated by the Jet Propulsion Laboratory, California Institute of
Technology, under contract with the National Aeronautics and Space
Administration.

This publication makes use of data products from the following
facilities: the Wide-field Infrared Survey Explorer, which is a joint
project of the University of California, Los Angeles, and the Jet
Propulsion Laboratory/California Institute of Technology, funded by
the National Aeronautics and Space Administration; the Spitzer Space
Telescope, which is operated by the Jet Propulsion Laboratory,
California Institute of Technology under a contract with NASA; the Two
Micron All Sky Survey, which is a joint project of the University of
Massachusetts and the Infrared Processing and Analysis
Center/California Institute of Technology, funded by the National
Aeronautics and Space Administration and the National Science
Foundation; the Digitized Sky Survey which was produced at the Space
Telescope Science Institute under U.S. Government grant NAG W-2166
using photographic data obtained by the Oschin Schmidt Telescope on
Palomar Mountain and the UK Schmidt Telescope; ISO, an ESA project
with instruments funded by ESA Member States (especially the PI
countries: France, Germany, the Netherlands and the United Kingdom)
and with the participation of ISAS and NASA; and the NRAO VLA Sky
Survey \citep{condon1998} which was performed by the US National Radio
Astronomy Observatory which is operated by Associated Universities,
Inc., under cooperative agreement with the National Science
Foundation.

\end{document}